\documentclass[fleqn,10pt]{wlscirep}
\usepackage[utf8]{inputenc}
\usepackage[T1]{fontenc}
\title{Transition Delay Using Biomimetic Fish Scale Arrays}

\author[1,*]{Muthukumar Muthuramalingam}
\author[2]{Dominik K. Puckert}
\author[2]{Ulrich Rist}
\author[1]{Christoph Bruecker}
\affil[1]{School of Mathematics, Computer Science and Engineering, City, University of London, London, United Kingdom EC1V 0HB}
\affil[2]{Institut für Aerodynamik und Gasdynamik, Universität Stuttgart, Pfaffenwaldring 21, 70569 Stuttgart, Germany }

\affil[*]{muthukumar.muthuramalingam@city.ac.uk}

\affil[+]{these authors contributed equally to this work}

\begin{abstract}
Aquatic animals have developed effective strategies to reduce their body drag over a long period of time. In this work, the influence of the scales of fish on the laminar-to-turbulent transition in the boundary layer is investigated. Arrays of biomimetic fish scales in typical overlapping arrangements are placed on a flat plate in a low-turbulence laminar water channel. Transition to turbulence is triggered by controlled excitation of a  Tollmien-Schlichting (TS) wave. It was found that the TS wave can be attenuated with  scales on the plate which generate streamwise streaks. As a consequence, the transition location was substantially delayed in the downstream direction by 55\% with respect to the uncontrolled reference case. This corresponds to a theoretical drag reduction of about 27\%. We thus hypothesize that fish scales can stabilize the laminar boundary layer and prevent it from early transition, reducing friction drag. This technique can possibly be used for bio-inspired surfaces as a laminar flow control means.
\end{abstract}
\begin{document}

\flushbottom
\maketitle
%
%
\thispagestyle{empty}

\section*{Introduction}
Fish are one of the oldest evolutionary species contributing to more than half of the living vertebrates distributed almost evenly across seawater and freshwater regions of the world\cite{ref3}$^,$\cite{ref4}. It comprises more than 33100 species and is larger than the sum of all other vertebrates, with the size range varying from a few millimetres to more than 10m. Of the 33100 species, more than 26000 belong to bony fish. Only 1000 species belong to cartilaginous fish, such as sharks, and just only 100 species belong to jawless fish (e.g. lamprey and hagfish)\cite{ref5}. Fish are highly dynamic creatures that persistently travel in water to reproduce, feed on their prey, and evade from their predator\cite{ref6}. As a result, much of the energy expended is largely for locomotion (against drag from skin friction and pressure drag) within the aquatic environment. This can be still water or turbulent water, which is often the case in river flows\cite{ref7}$^,$\cite{ref8}.

Fish have prodigious features for flow control over their bodies, adapted to the environment and living circumstances. The skin and appendages are essentially the main parts where the flow is most likely to be tweaked to meet the need. For example, sharks swim at speeds ranging from 0.3 to 0.9 body-length/sec, reaching Reynolds numbers\footnote{https://en.wikipedia.org/wiki/Reynolds\_number} more than a million, which makes the boundary layer turbulent over most of their body\cite{ref9}. The placoid scales (similar to a riblet shape) have been proven to reduce the turbulent skin friction drag\cite{ref10}. Most studies over a decade focused on this specific scale shape\cite{ref11}$^,$\cite{ref12}. Similarly, pectoral flippers on humpback whales have leading-edge tubercles that prevent stalling and give them high manoeuvrability\cite{ref1}. Dolphins (aquatic mammals) reduce their drag by delaying the transition to turbulent flow on their body due to their anisotropic and compliant skin structure which dampens the flow instabilities\cite{ref13}.

This study will focus on the hydrodynamics of the skin of bony fish. More than 95\% of the existing bony fish belong to teleosts whose skin is covered by leptoid scales that are further classified into scales of ctenoid and cycloid type\cite{ref5}. Scale classification and morphology from various fishes were determined and inferred for possible hydrodynamic functions\cite{ref19}$^,$\cite{ref20}.  Most teleost fish operate at the Reynolds number range where transitional boundary layer flow prevails on the fish surface\cite{ref9}$^,$\cite{ref15}. In addition, their elongated body with an elliptical cross-section  resembles that of an hydrofoil. Therefore, flow over these bodies can be closely related to the flow over a flat plate. Very few research on the flow dynamics over typical skins of teleost have been performed so far. For grass carp,  (\textit{Ctenopharyngodon idellus}) some geometric parameters of the scales were scanned and a bionic surface  was created with individual, non-overlapping elements resembling those scales. When tested on a flat plate in a towing tank, the results showed  a drag reduction of approximately 3\%\cite{ref14}. Recently, some of the present authors have investigated the scale structure of European sea bass (\textit{Dicentrarchus labrax}) and designed a biomimetic surface, which mimics the realistic features of overlapping scales and their characteristic surface pattern. Computation Fluid Dynamics (CFD) was used to study the flow pattern over the surface and revealed a hitherto unknown effect of the scales as a mechanism to  generate a regular pattern of parallel streamwise velocity streaks in the boundary layer\cite{ref2}. To prove their existence also on the real fish skin, oil flow visualisation was done on sea bass and common carp, which indeed confirmed their presence in a regular manner along their real body, with the same arrangement relative to the scale array as observed along the biomimetic surface. These results let the authors hypothesize about a possible mechanism for transition delay, inspired by various previous fundamental transition studies, where streaky structures generated by cylindrical roughness elements or vortex generator arrays have shown a delay of transition \cite{ref16}$^,$\cite{ref17}$^,$\cite{ref18}.  

\begin{figure}[hbt!]
\centering
  \includegraphics[width=0.69\textwidth]{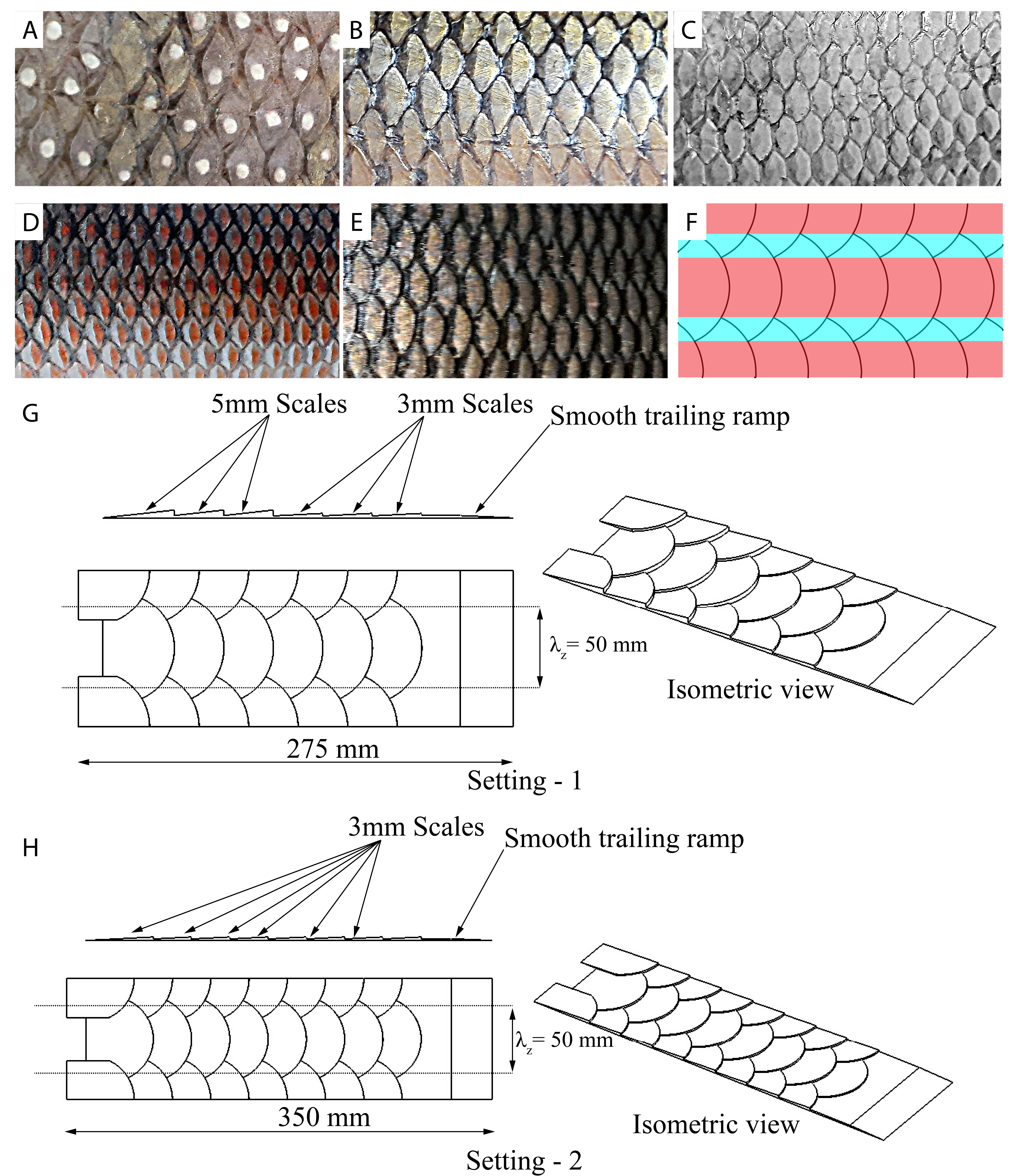}
\caption{\textbf{Scale structure on different fishes, biomimetic scale array and CAD drawings of fish scale array} (A) Etroplus (\textit{Etroplus suratensis}). (B) Mrigal carp (\textit{Cirrhinus cirrhosus}). (C) Tilapia  (\textit{Oreochromis niloticus}). (D) Rohu (\textit{Labeo rohita}). (E) Catla (\textit{Labeo catla}). (F) Biomimetic scale structure (Pink: Central region, Light blue: Overlap region. (G) Setting-1 which has first three rows with 5mm thick scales followed by three rows of 3mm thick scales. (H) Setting-2 which has eight rows of 3mm thick scales.}
\label{fig:figure1}      
\end{figure}

The main purpose of this study is to prove this hypothesis with experiments in a well-established low-turbulence water channel facility, where the transition process can be studied under well-defined boundary conditions with controlled excitation of the fundamental instability of the Tollmien Schlichting waves. In addition, the facility allows optical access to the boundary layer flow and detailed measurements of the instability using hot-wire. The facility is established as a unique low-turbulence laminar water channel and there has been a long history of transition experiments\cite{ref27}$^,$\cite{ref30}$^,$\cite{ref31}. 
%
%
%
%
%
The biomimetic scale array used in this study is based on the scale structure of European sea bass (\textit{Dicentrarchus labrax}), and its details have been published in the authors' preceding paper\cite{ref2}. Although the geometry was derived from sea bass, the scale pattern is common to most species of seawater and freshwater fishes. Figure.\ref{fig:figure1}F displays the scale pattern used in this study where the area with pink colour is the central region and the area with a light blue colour represents the region of overlap between adjacent scales. In a direct comparison with real fish scales, the biomimetic scale pattern in Fig.\ref{fig:figure1}F looks very similar to the natural scales of the species in Fig.\ref{fig:figure1}A-E.

\begin{figure}[hbt!]
\centering
  \includegraphics[width=0.6\textwidth]{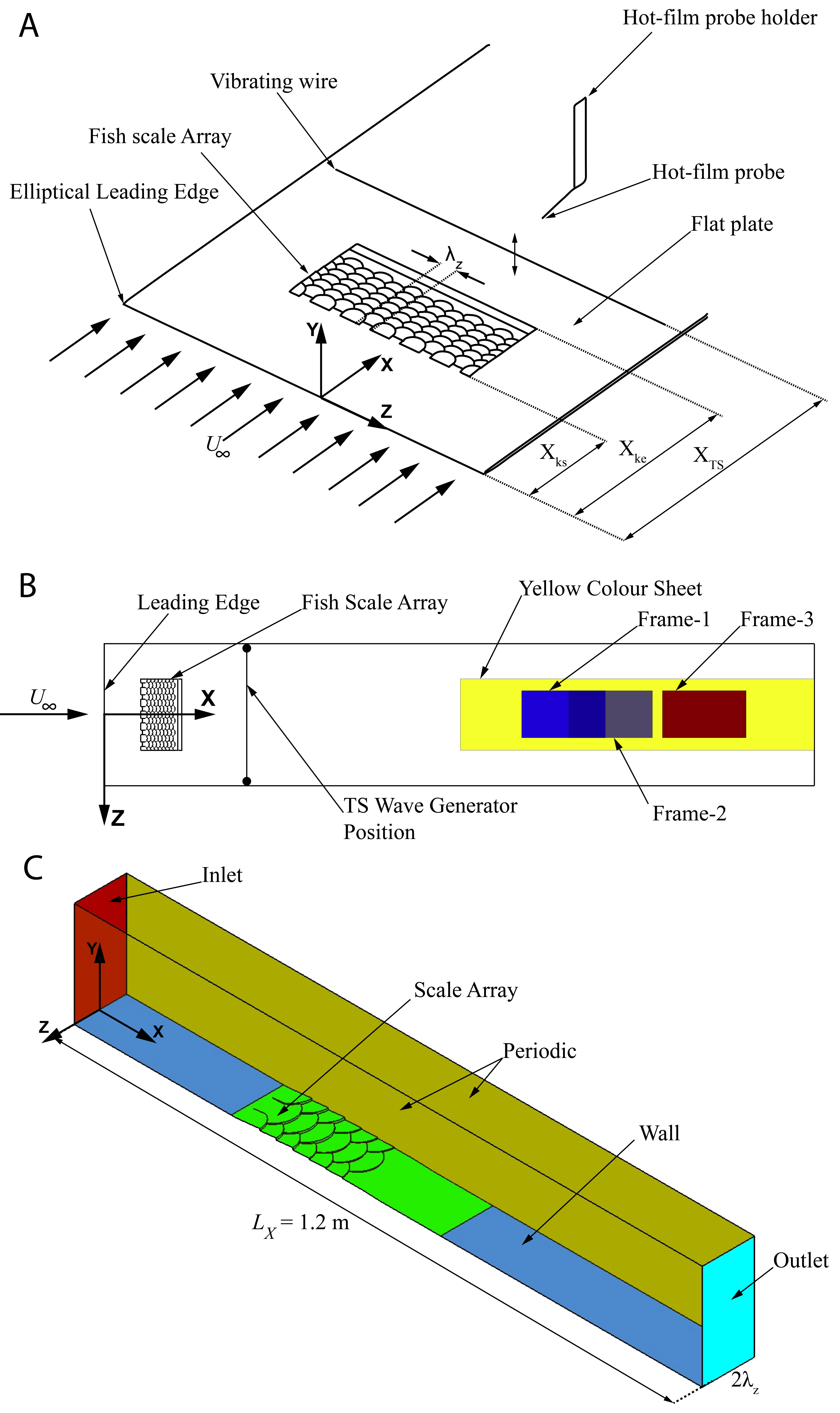}
\caption{\textbf{Experimental flow set-up, details of flow visualisation set-up and computational domain} (A) Experimental Set-up in the Laminar Water channel with a biomimetic scale array. (B) Experimental arrangement from top view, Yellow area is the region in which a yellow sheet was glued on the flat plate. All three frames where videos were recorded are shown by rectangles. Note that Frame-1 and Frame-2 have some overlap. (C) CFD domain with periodic conditions. Setting-1 is shown in this figure and for Setting-2 all conditions remain the same except the fish scale array geometry. The inlet was provided with velocity parallel to $X$-axis with a value of $U_\infty = 0.086$m/sec.}
\label{fig:figure2}      
\end{figure}

\section*{Experimental Conditions}
The experiments were conducted in the Laminar Water Channel of the Institute of Aerodynamics and Gas Dynamics in Stuttgart, Germany. The test section dimensions are 10m in length (streamwise direction: $X$-axis), 1.2m in width (spanwise direction: $Z$-axis) and 0.2m in height (wall-normal direction: $Y$-axis) with a turbulence intensity value less than 0.05\%\cite{ref18}. Hot-film anemometry was used to measure the velocity of the flow inside the water channel. A flat plate was used to generate a laminar boundary-layer flow over it at a flow velocity ($U_\infty$) of 0.086m/sec corresponding to a Reynolds number of $Re_L \sim 5.2 \cdot 10^5$ based on the flat plate length ($x = L$) of 6m, where the Reynolds number is defined by ($Re_x = xU_\infty/\nu$) and $\nu$ is the kinematic viscosity of water. The experimental details are shown in Fig.\ref{fig:figure2}A. The leading edge of fish scale arrays was located at ($X_{ke}$) 0.3m from the leading edge and two models were used to develop different streak amplitudes. The first model, Setting-1 was 3D printed with three rows of scales with 5mm height followed by three rows of scales with 3mm height. The second model, Setting-2 was 3D printed with eight rows of scales with 3mm height. The trailing edge of the model has a smooth ramp to avoid separation behind the model. The roughness height of the scales ($h_s$) on sea bass was around 0.15mm for a fish length of 350mm. At a swimming speed of around 1.5m/sec the boundary layer thickness ($\delta$) will be about 0.5mm at the location where the scale array begins, which leads to a $\delta/h_s$ ratio of 3. Hence, in this study the $\delta/h_s$ ratio is maintained to match the dynamic similarity after the upscaling of the biomimetic scale array\cite{ref34}. This is a common practice in fluid mechanics to upscale the model based on geometric and kinematic similarities, for example, see the tests on upscaled shark skin scales here\cite{ref37}. The CAD drawings of both models are shown in Fig.\ref{fig:figure1}G and Fig.\ref{fig:figure1}H in detail. The roughness Reynolds number defined by  $R_{kk} = ku_k/\nu$ is 334 for Setting-1 and 143 for Setting-2, where $u_k$ is the undisturbed velocity without roughness at the height ($k$) of the maximum roughness\cite{ref24}. Both models were 3D printed in spanwise segments of 0.2m so that the length of the total setting in $Z$-axis was 0.6m. A standard procedure in transition research along a flat plate is used herein to investigate the response of the boundary layer to the modified surface. The method uses a vibrating wire, which is spanned in the spanwise direction and used to excite a 2D Tollmien-Schlitching wave\cite{ref21}. Location of the wire within the boundary layer and the vibration frequency are adapted to the theoretical instability diagram of the Blasius solution for a laminar 2D boundary layer\cite{ref21}.  A wire of 0.1mm diameter is located at $X_{TS}$ = 1.2m from the leading edge of the flat plate\cite{ref18} at a wall-normal distance of 5mm from its surface and vibrating at a physical frequency of $f$ = 0.2Hz (the corresponding normalised frequency is $F = 2 \pi f\nu /U_\infty^2 = 166 \cdot 10^{-6}$) with an amplitude of 0.25mm.

\begin{figure}[ht]
\centering
  \includegraphics[width=0.8\textwidth]{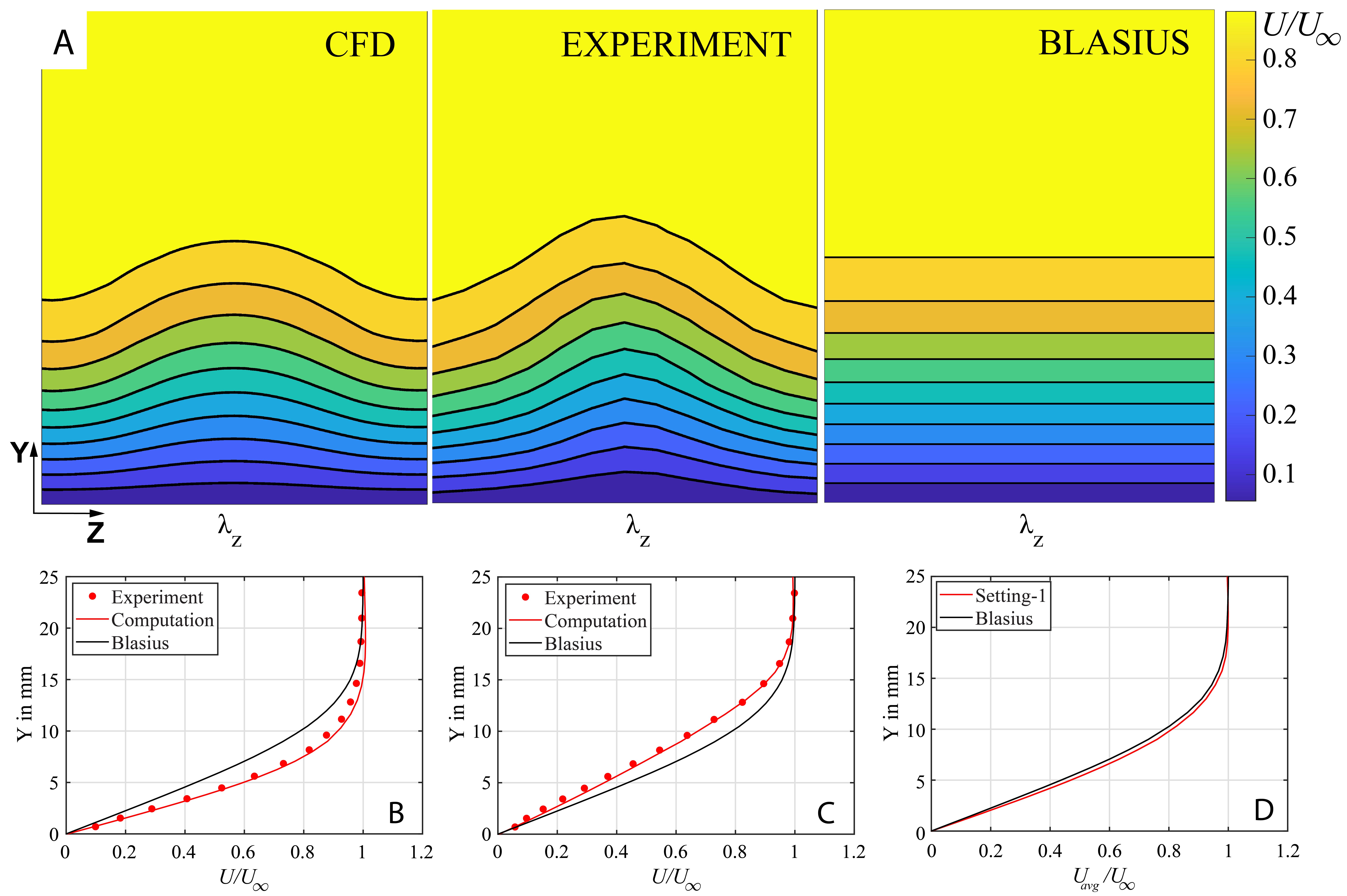}
\caption{\textbf{Comparison of CFD with experimental results} (A) Velocity variation of streaky base flow in $Z-Y$ plane at a distance of 1.2m from the leading edge (Left - CFD result, Middle - Hot-film result, Right - Blasius). Normalised velocity profile at (B) the overlap region (high-velocity region) behind the scale array $Z = 0.025$m. (C) at the central region (low-velocity region)  behind the scale array $Z = 0$m. (D) Spanwise averaged streamwise velocity profile for Setting-1. Black line - Reference Blasius velocity profile}
\label{fig:figure3}       
\end{figure}
\pagebreak

\section*{Results}
The surface, mimicking the array of overlapping scales along the flat plate in the low-turbulence facility, again show the generation of streamwise velocity streaks, similar as detected in our previous study on real fish bodies\cite{ref2}. Contours of constant streamwise velocity in a cross-section of the boundary layer  in Fig.\ref{fig:figure3}A illustrate the velocity variation at 1.2m from CFD simulation and experiment for one wavelength of the streak ($\lambda_z$ = 50mm) for Setting-1. Note, that the result is periodic in spanwise direction with each row of scales. The Blasius contour at the right depicts the 2-D mean flow over the same wavelength\cite{ref32}. A detailed comparison of velocity profiles in the high- and low-velocity regions behind the scales are shown in Fig.\ref{fig:figure3}B and Fig.\ref{fig:figure3}C, respectively. In the region of high velocity and low velocity regions the deviation from the reference Blasius profile depicts velocity deficit or increase behind the scale array. Spanwise averaged streamwise velocity in the streaky flow is compared with Blasius profile in Fig.\ref{fig:figure3}D. The streaky flow produces a fuller velocity profile when compared with the reference flow which results in a shape factor of 2.47 instead of 2.59 which is comparable with the Large Eddy Simulation results reported in literature for the streaky base flow\cite{ref35}. CFD and experimental results are comparable with only minor variations that may be attributed to the boundary conditions and the experimental uncertainties. Measurements at different locations further downstream prove that the streak persists in streamwise direction (not shown here).   

\begin{figure}[ht]
\centering
  \includegraphics[width=0.6\textwidth]{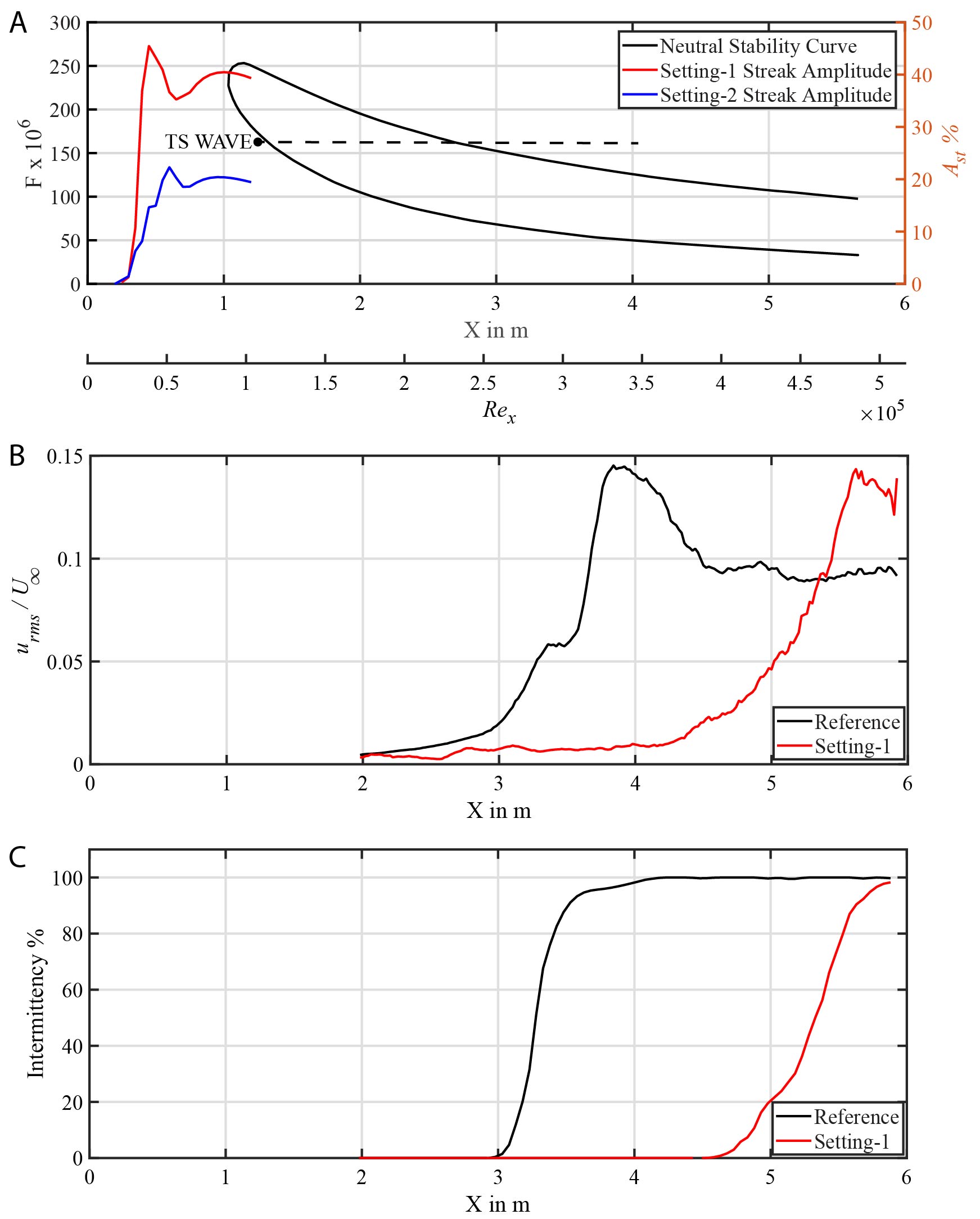}
\caption{\textbf{Neutral stability curve, root mean square of velocity fluctuations and intermittency variation} (A) Neutral stability curve and streak amplitude for two settings. (Normalised frequency $F$ given in left 'y' coordinate and streak amplitude $A_{st}$ given in right 'y' coordinate, Black dot marks the TS wave generator position and dotted line indicates the frequency value). (B) $u_{rms}$ plot normalised by free stream velocity ($U_\infty$) for reference flat plate and Setting-1. (C) Intermittency plot for reference flat plate and Setting-1}
\label{fig:figure4}      
\end{figure}

This modulation of the velocity is fundamentally different from the streaky structure generated by the lift-up effect caused by a vortex generator or cylinder array\cite{ref23}. The streaky structure generated by the overlapping scales is formed by a spanwise periodic flow very close to the surface, and the amplitude of the streak increases with the number of scale rows in the direction of flow. Setting-1 with about the same number of scales as Setting-2 but twice as thick produces a streak amplitude approximately twice as large compared to Setting-2 ($A_{st} \sim 40\%$ for Setting-1  and   $A_{st} \sim 20\%$ for Setting-2)  as shown in Fig.\ref{fig:figure4}A. The streak amplitude is defined as in Eqn.\ref{eq:1} and increases with the number of scale rows from the leading edge of the model, it is seen from Fig.\ref{fig:figure4}A within X = 0.3m to X = 0.6m (It is the extent where the scales are placed on the tunnel).  Afterwards it drops  as a result of the decelerating trailing ramp. Once again the flow reorganises up to some extent to increase the streak amplitude and then the viscosity causes it to decay continuously downstream. Both models did not induce a bypass transition (instantly tripping laminar flow into turbulent), nor did they induce secondary streak instability as seen from hot-film signals and with flow visualisations.
   \begin{equation}\label{eq:1}
   A_{st} = \bigg[\max_{y}\{U(X,y,z)\}-\min_{y}\{U(X,y,z)\}\bigg]\bigg/({2U_\infty})
   \end{equation}
%

To investigate the response of the boundary layer to the scaled surface with regard to the laminar-to-turbulent transition process, a controlled transition experiment with a representative Tollmien-Schlichting (TS) wave  at a given frequency  were performed, following the method invented in \cite{ref21}. In Fig.\ref{fig:figure4}A the neutral stability curve is shown as a black line for the present free-stream velocity of $U_\infty = 0.086$m/s. The neutral stability curve is given along the $X$-axis and also for comparison with non-dimensional parameters in similar studies based on the Reynolds number ($Re_x$). The area within the stability curve is the region in which, according to linear stability theory, infinitesimal disturbances will grow exponentially\cite{ref22}.
The velocity signals were measured for 60sec at a data acquisition frequency of 100Hz at $Y$ = 10mm from the wall from $X$ = 1.98 m to 5.92m for the reference flat plate and fish scale array (Setting-1) cases.  In the reference case, the induced small disturbances from the vibrating wire grow in the streamwise direction inside the instability region which can be inferred from the black $u_{rms}$  curve from 2m to 2.6m in Fig.\ref{fig:figure4}B. Initially, the so-called primary instability mechanism increases the velocity fluctuations until secondary instability mechanism set in, afterwards the fluctuations increase rapidly until they reach a peak around 3.8m in  Fig.\ref{fig:figure4}B. From there on the flow is turbulent as observed from the constant $u_{rms}$ plateau after 4.5m\cite{ref25}. However, for the flow with fish scale array (Setting-1), as seen from the red line in Fig.\ref{fig:figure4}B the fluctuation level $u_{rms}$ remains almost constant until 4m and it increases monotonically but with a lower rate when compared with the reference case.
The local flow state can be defined generally by the intermittency parameter which classifies the flow into laminar, turbulent, and transitional \cite{ref26}. The value in percentage indicates the nature of the flow, e.g.: zero represents laminar flow and 100\% represents fully turbulent flow, and any value in between indicates how long the flow is turbulent in a given period of time. For the reference flat plate case, the flow is laminar until 3m and becomes turbulent just after 4m as shown in the black curve in Fig.\ref{fig:figure4}C. For the case with fish scale array as shown in the red line in Fig.\ref{fig:figure4}C, the flow remains laminar for a larger extent until 4.65m and then becomes turbulent around 6m. This results in a streamwise extension of laminar flow by about 1.65m which corresponds to a 55\% delay in transition location.

\begin{figure}[ht]
\centering
  \includegraphics[width=0.832\textwidth]{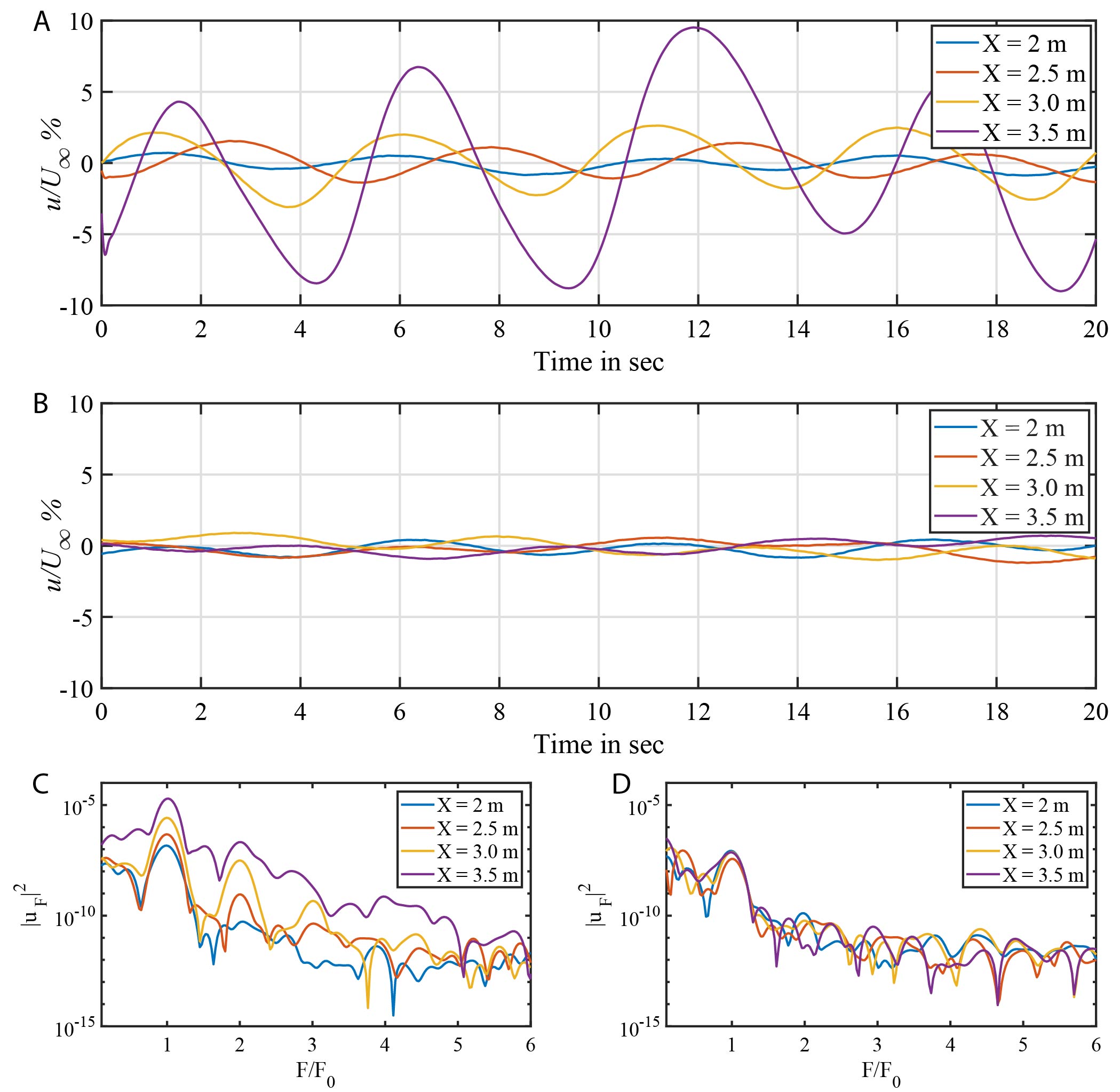}
\caption{\textbf{Instantaneous velocity and spectrum for reference flat plate and fish scale array case} (A) Instantaneous velocity at four locations for reference flat plate case. (B) Instantaneous velocity at four locations for fish scale array case (Setting-1). (C) Spectrum for the velocity signals for reference flat plate case.  (D) Spectrum for the velocity signals for fish scale array case (Setting-1).}
\label{fig:figure5}      
\end{figure}

\begin{figure}[ht]
\centering
  \includegraphics[width=0.8\textwidth]{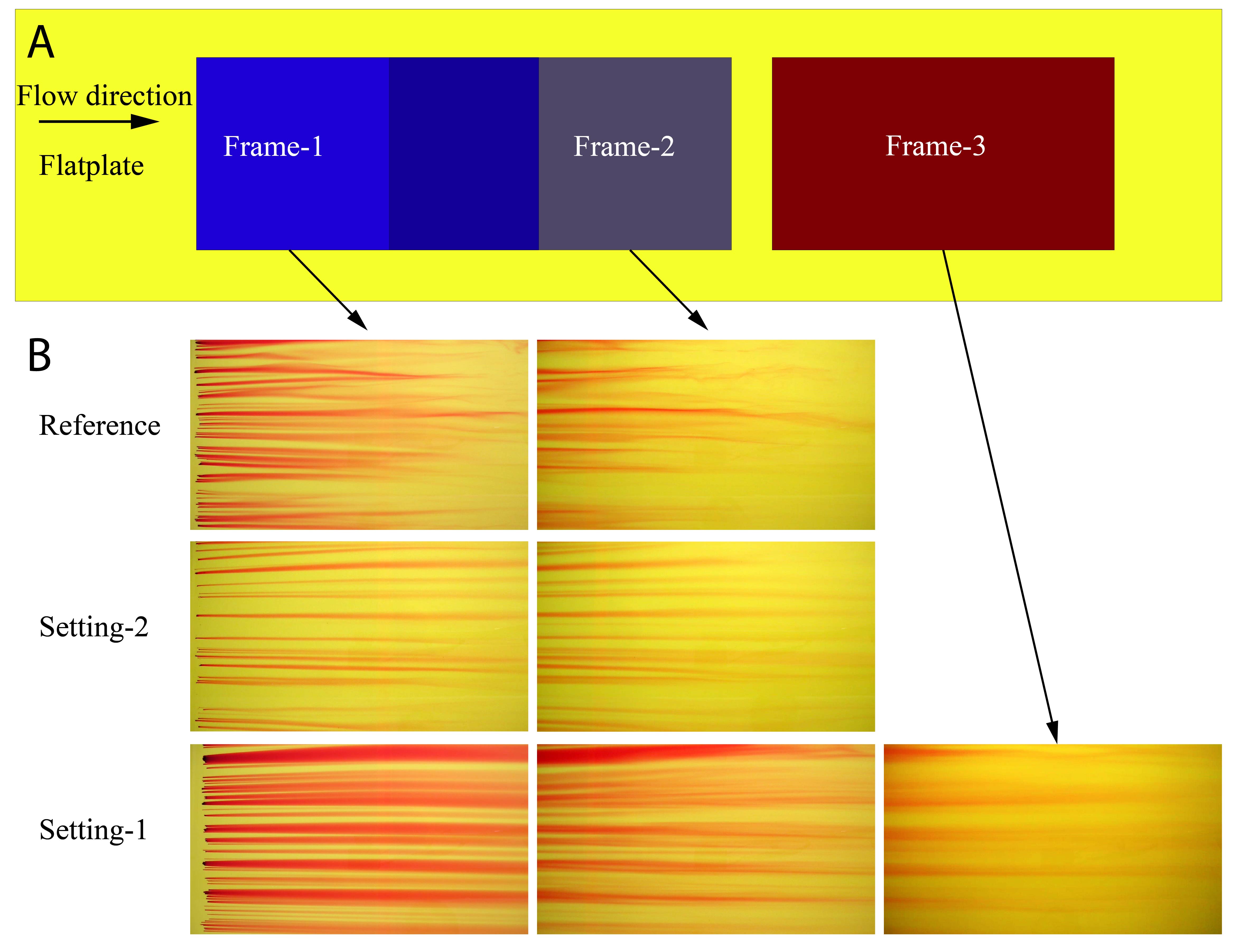}
\caption{\textbf{Flow visualisation results} (A) Flow visualisation setting, Frame-1: 3.15 to 3.85m, Frame-2: 3.55 to 4.25m, Frame-3: 4.35 to 5.05m. (B) Flow visualisation pictures for reference, Setting-2 and Setting-1 cases.}
\label{fig:figure6}      
\end{figure}

\begin{figure}[ht]
\centering
  \includegraphics[width=0.85\textwidth]{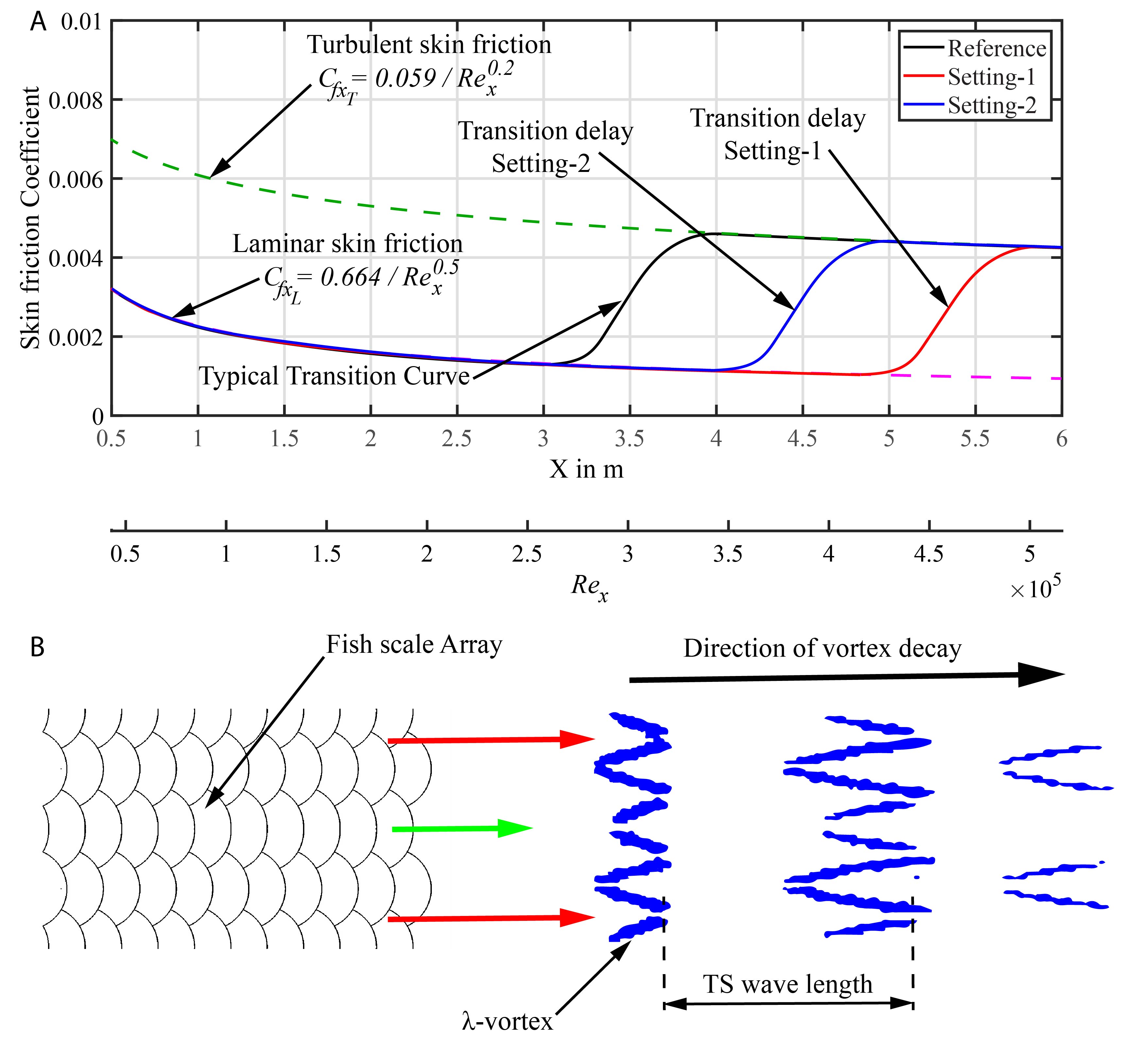}
\caption{\textbf{Typical skin friction plot and proposed transition delay mechanism} (A) Skin friction coefficient curve: Dashed green line - Turbulent flow, Dashed pink line - laminar flow. (B) Flow behind fish scale array with modulated TS wave. Red arrow - High speed region, Green arrow - Low speed region.}
\label{fig:figure7}      
\end{figure}

Next, we explore the flow by means of temporal velocity signals for the two cases previously discussed. Figure.\ref{fig:figure5}A displays the velocity signals subtracted from their mean values at 2, 2.5, 3.0 and 3.5 m for a period of 20 sec for the reference case without fish scales. The amplitude of the velocity signals in Fig.\ref{fig:figure5}A increases with streamwise coordinate $X$. The respective frequency spectra of these velocity signals is given in Fig.\ref{fig:figure5}C, where the abscissa is normalised with respect to the vibration frequency ($F_0$ = 0.2Hz) of the vibrating ribbon. At $X=$2m the spectrum displays a peak at $F/F_0$=1 which indicates that the fluctuation energy is only from the wire's fundamental vibrating frequency. Higher harmonics and subharmonics of the fundamental frequency $F_0$ appear further downstream (see the peaks at 3 and 3.5 m) and the disturbance energy also increases compared to the spectrum at 2m. At 3.5m the energy is increased over all frequencies given in the plot indicating that the flow is becoming increasingly disturbed resulting in turbulence. On the contrary, the velocity magnitude for the flow with fish scale array remains within 2\% at all locations, as shown in Fig.\ref{fig:figure5}B.  At the same time the fluctuation energy is very small compared to the flat plate case as shown in Fig.\ref{fig:figure5}D. Additionally, the higher harmonic components in the flow are completely absent in the case of the fish scale array. This reflects the very low level
of velocity fluctuations $u_{rms}$  depicted by a red line in in Fig.\ref{fig:figure4}B. The increase of $u_{rms}$ beyond 4m is due to uncontrolled background oscillations of the water tunnel and not necessarily due to a re-amplification of the TS wave.

Complementing information about the flow states in different cases is obtained from flow visualisation using the method of surface streakline generation. Individual potassium permanganate crystals were placed on the flat plate and, while dissolving as dye in the water, they visualize the flow close to the surface. These streak lines will be visible as compact dye lines if the flow is laminar, while, in contrast, they develop kinks and diffuse very quickly when the flow is turbulent. The locations where the visualisations has been done is shown in Fig.\ref{fig:figure6}A for brevity. Figure.\ref{fig:figure6}B(Reference) shows the visualisation picture for the reference flat plate case from $X=$3.15 to 3.85m. Certainly, the streaklines are visible for more than 70\% of the picture indicating laminar flow with some instabilities. In  Figure.\ref{fig:figure6}B(Reference), where the frame is from 3.55 to 4.25m the streaklines have broken down because of the turbulent flow which is comparable with the hot-film measurements explained above. Figure.\ref{fig:figure6}B(Setting-1) portray the visualization picture for the same locations described previously, but for the case with fish scale array (Setting-1). The flow is completely laminar in the given locations and even for an additional location from 4.35 to 5.05m as shown in Fig.\ref{fig:figure6}B(Setting-1). For the case with the second set of fish scale array (Setting-2), the flow is still laminar in the above-mentioned locations until the end of the picture as shown in Fig.\ref{fig:figure6}B(Setting-2). Hence, this visually proves that the fish scale array increases the laminar flow extent by delaying the transition and this result is in perfect agreement with the transition delay visualisations performed with cylindrical roughness elements\cite{ref17}. 

\subsection*{Drag estimation} 
Flow over any body will experience drag that has two components, skin friction and pressure drag. For a flat plate, the drag is only from skin friction with the friction coefficient for laminar and turbulent flow given by Eqn.\ref{eq:2} and Eqn.\ref{eq:3}, respectively\cite{ref32}. These equations were compared with Direct Numerical Simulation results and found to be comparable with similar kind of TS waves\cite{ref33}.
     \begin{equation}\label{eq:2}
     C_{fx_L} = 0.664/\sqrt{Re_x}
     \end{equation}
     
     \begin{equation}\label{eq:3}
    C_{fx_T} = 0.059/{Re_x}^{\frac{1}{5}}
     \end{equation}

 Hence, the total drag along a flat plate with laminar and  turbulent flow regimes can be approximated by the summation of the drag components by the Eqn.\ref{eq:4}, where, $x_L$ is the location from the leading edge of the flat plate where the flow is assumed to change from laminar to turbulent state. 

     \begin{equation}\label{eq:4}
     D_{Net} = D_L + D_T + D_P = \int_{0}^{x_L} \frac{1}{2}C_{fx_L} \rho U_\infty^2 dx + \int_{x_L}^{L} \frac{1}{2}C_{fx_T} \rho U_\infty^2 dx + D_P
     \end{equation}
For the sake of comparing the drag for both cases, the location $x_L$ is assumed where the intermittency factor reaches a value of 50\%. For the reference flat plate case the location $x_L$ is estimated from the hot-film measurements to be at 3.3m and for the case with the fish scale array (Setting-1) the location $x_L$ is placed at 5.3m. For Setting-2 the location $x_L$ is chosen from the flow visualisation of about 4.3m since we do not observe any turbulence even at Frame-2. Additionally, when a body like fish scale array is mounted on the flat plate it experiences added pressure drag ($D_P$). The value of this drag component is calculated from the CFD simulation for a length of $L_X$ = 1.2m (as shown in Fig.\ref{fig:figure2}C). The components of drag for the two cases are given in Table.\ref{table:1}. The net drag (D\textsubscript{Net}) is reduced by 0.02N with the fish scale array (Setting-1) and 0.008N for Setting-2. This results in a net drag reduction of about 27\% for Setting-1 and 10.7\% for Setting-2 when compared with the reference flat plate case.\\ 
\begin{table}[ht]
\begin{tabular}{cccccc}
\hline
\multicolumn{1}{|c|}{Configuration} & \multicolumn{1}{c|}{Laminar region x$_L$ (m)} & \multicolumn{1}{c|}{Laminar drag D$_L$ (N)} & \multicolumn{1}{c|}{Turbulent drag D$_T$ (N)} & \multicolumn{1}{c|}{Pressure drag D$_P$ (N)} & \multicolumn{1}{c|}{Net drag D$_{Net}$ (N)} \\ \hline
\multicolumn{1}{|c|}{Reference}    & \multicolumn{1}{c|}{3.3}                   & \multicolumn{1}{c|}{0.030421}         & \multicolumn{1}{c|}{0.044954}           & \multicolumn{1}{c|}{0}                 & \multicolumn{1}{c|}{0.075375}     \\ \hline
\multicolumn{1}{|c|}{Setting - 1}   & \multicolumn{1}{c|}{5.3}                   & \multicolumn{1}{c|}{0.038553}         & \multicolumn{1}{c|}{0.011172}           & \multicolumn{1}{c|}{0.005648}          & \multicolumn{1}{c|}{0.055373}     \\ \hline
\multicolumn{1}{|c|}{Setting - 2}   & \multicolumn{1}{c|}{4.3}                   & \multicolumn{1}{c|}{0.034726}         & \multicolumn{1}{c|}{0.027573}           & \multicolumn{1}{c|}{0.005048}          & \multicolumn{1}{c|}{0.067347}     \\ \hline
\multicolumn{1}{l}{}                & \multicolumn{1}{l}{}                       & \multicolumn{1}{l}{}                  & \multicolumn{1}{l}{}                    & \multicolumn{1}{l}{}                   & \multicolumn{1}{l}{}             
\end{tabular}
\caption{Comparison of estimated drag for reference flat plate and fish scale array (Setting-1)}
\label{table:1}
\end{table}

This result can be understood by using the skin friction plots as depicted in Fig.\ref{fig:figure7}A. The laminar skin friction curve is shown as a dashed pink line and the turbulent skin friction curve as a dashed green line. Typical transition curves appear in all the cases considered here\cite{ref32}. Generally, if a flow becomes turbulent the skin friction coefficient rises to almost twice its value for laminar flow at a particular location. The total drag of the surface is the area under the skin friction curve, therefore, the area under the curve reduces for fish scale array when compared with the reference flat plate case. Furthermore, the reduction in integral amplitude is proportional to the streak amplitude in the experiments considered here.

\section*{Conclusions}

In this paper, we have investigated and shown the stabilizing influence of a periodic streaky base flow generated by a biomimetic scale array along a flat plate boundary layer
on a Tollmien Schlitching wave. In the absence of the fish scale array (reference Blasius case) the TS wave grows inside the boundary layer as a two-dimensional fluctuation. When the amplitude increases three-dimensional undulations appear which lead to $\lambda$-vortices. These vortices will increase the three-dimensional velocity fluctuation until they form staggered or aligned vortices based on H-type or K-type transition\cite{ref35}. These structures will further develop into hairpin eddies in the fully turbulent region to increase the skin friction value. However, with the streaky base flow, the spanwise averaged boundary layer has a lower shape factor value in the laminar boundary layer region suggesting that the flow is stable as a whole. This change from the Blasius shape factor value of 2.59 depends on the amplitude of the streaks. The streaky base flow reorganises the TS wave, based on the velocity variation in the spanwise direction as depicted in Fig.\ref{fig:figure7}B. The high-speed region pushes a portion of TS wave to a larger extent in streamwise direction when compared with the low speed region to form a $\lambda$ vortex. However, because of the velocity variation in the spanwise direction the $\lambda$ vortices decay with length. This proposed transition delay mechanism is based on the CFD results which is very similar to the results from the Large Eddy simulations\cite{ref35}.

As explained in the introduction, most fish species operate in the transitional laminar-to-turbulent Reynolds number regime which was tested in this study (i.e., $Re_x \sim 10^5$). Experiments revealed that the scale array attenuates the TS wave and hence is able to delay laminar-turbulent transition which results in a maximum net drag reduction of about 27\% for the given configuration. The present mechanism to generate the streaks differs from those in previous studies where cylindrical roughness elements and vortex generators were used to delay transition. Cylinder arrays can produce maximum streak amplitudes of about 12\%, beyond that, the cylinders will trigger bypass transition because of the absolute instability in the wake\cite{ref25}. In addition, the roughness elements generate parasitic drag due to the pressure drop around the protruding body. Vortex generators can delay transition with higher streak amplitudes without secondary instabilities, however, they will act as bluff bodies when the flow is not perfectly aligned with the orientation of the vortex generator. Additionally, in the case of purely laminar flow where transition does not occur (i.e. $Re_x < 10 ^ 4$), these two types of bodies will always generate parasitic drag. 

 To be effective at all operating Reynolds numbers, it requires a multi-role flow control mechanism. Our previous studies showed that the biomimetic fish scale array already achieved laminar drag reduction in a clean laminar flow (where the ratio of boundary layer thickness to the maximum scale height was greater than 10), thus reducing the drag in the low-velocity regime, too\cite{ref2}. The mechanism by which the overlapping scales generate the streaks is via producing a spanwise flow near the wall, which is sustained by the repeated overlapping along the rows of the scales. When the swimming speed increases and the flow is likely to be transitional, the streaks from the scales tend to prolong the laminar flow by delaying the transition without adding any parasitic drag, thus minimizing skin friction. Note, that the results further indicate no tendency of the scales to generate a by-pass transition. Therefore this mechanism is assumed to be robust against moderate variations in details, like scale shape and height. The results let us speculate that the overlapping scale arrays on most bony fishes are an evolutionary result to minimize friction drag by producing streaky flow which produces a fuller laminar velocity profile on the surface.
 
 Despite the promising results from this study, salient limitations should be mentioned. Primarily, the tests were done on a flat plate ignoring the pressure gradient which is inevitable on the body of the fish. However, recent experiments on an aerofoil with streaky base flow delays transition for a particular configuration which motivates the use of these fish scale array on a surface with imposed pressure gradient\cite{ref36}. Secondly, the flexibility of the scales and also the undulatory motion of the fish is not considered here, which largely changes the transitional boundary layer\cite{ref29} due to unsteady effects which were not part of the present study. Therefore, the primary focus was on the performance of the biomimetic fish scale array on a very controlled transition scenario for comparison with previous studies which were successful in delaying flat-plate boundary-layer transition. From the observation that streaks exist, already proven with the help of surface flow visualisations on a real fish, and the observation that scale arrays delay transition without by-pass transition, it is hypothesized that the fish scales are efficient in delaying laminar-turbulent transition on a real fish body, as well. Additionally, the performance of the scale array (rigid or flexible) in turbulent or separated boundary layers remains to be studied in future research. \\

\section*{Methods}
\subsection*{Biomimetic fish scale models}
Biomimetic fish scale models were 3D printed at City University of London and University of Stuttgart using ABS plastic material of density about 1080kg/m\textsuperscript{3}. The array was made in many pieces because of the size restriction in the printer. The models were modeled in CATIA and were 200mm wide and printed as separate tiles with extra 1mm thickness at the base for stable print. The leading part with smooth ramp and three scale rows was printed as one piece and the other three scale rows with trailing edge ramp were printed as a separate piece. When placed one behind the other it will be acting as a single array of scale rows. The tile with scale array was glued on the flat plate of the Laminar Water Channel to keep it in place without any movements. Two models have been used in this study as shown in Fig.\ref{fig:figure1}G and Fig.\ref{fig:figure1}H. 

\subsection*{Experiments using water channel}
 The experiments were conducted in the open channel closed loop Laminar Water Channel (Laminarwasserkanal) at University of Stuttgart, Germany. The flow is induced by two axial propellers connected to a frequency control drive to vary the RPM. The flow from the pump is passed through a honeycomb chamber followed by a long diffuser with a series of screens to a settling chamber with a contraction ratio of 7.7:1. Before the contraction three additional sets of screens were used to reduce the turbulence. The turbulence intensity lies below 0.05\% within a frequency range of 0.1 - 10Hz at 0.145m/s\cite{ref27}. The dimensions of the test section are 10m in length, 1.2m in width and 0.2m in height. Inside the test section a very long but segmented glass plate is used to create a two dimensional Blasius boundary layer. The leading edge of the first plate is elliptical to reduce the leading edge separation and getting zero pressure gradient on the flat plate quickly. In the spanwise direction the plates are little less than the widths of the water channel which provides natural suction to prevent the corner flows. Constant temperature hot-film anemometry was used to measure the velocity of the water using DANTEC  55R15 with a 16-bit A/D converter. The overheat ratio for the hot-film probe was set at 8\% as given in the manufacturer manual. All the measurements in this study have been acquired at 100Hz for 60 sec. Before starting the measurement the hot-film was calibrated in still water by traversing the probe with a controlled series of constant velocities to acquire the corresponding voltage from the data acquisition system. Finally a correlation graph was used to find the coefficients in King's law to relate the voltage (E) with velocity (U) given by 
   \begin{equation}
  U = \left[ \frac{E^2 - A}{B} \right]^\frac{1}{C} .
   \end{equation}
The data was post-processed in MATLAB for filtering the signals and also to find the spectrum of the signals. For details on the experimental facility and measurement equipment, the reader is referred to\cite{ref28}. 

\subsection*{Flow visualisation}
To perform flow visualisations the water channel was emptied below the flat plate to dry the surface. A yellow colour sheet of width 600mm and length of about 2500mm was placed on the flat plate for better contrast. Then it was left to glue on the surface for a day. The yellow colour sheet was placed from 3 to 5.5m on the flat plate where the flow disturbances grow from laminar to turbulent flow. Potassium permanganate crystals (less than 2mm in average) were placed at the start of the yellow sheet. As the crystals dissolve with water they will colour the water without changing its physical properties. The coloured water will be clearly identified in the regions where the flow is laminar and quickly diffuse and disappear in the turbulent flow regions. The dye flow visualisation was recorded from above the tunnel using digital cameras from three regions. All the video recordings were done with a shutter time of 1/30 seconds. The first frame was from 3.15 to 3.85m, the second frame  from 3.55 to 4.25m and the third frame  from 4.35 to 5.05m. The setup of the flow visualisation is shown in Fig.\ref{fig:figure2}B.\\

\subsection*{CFD Methodology}
The computational study was done using ANSYS-Fluent 19.0. The CFD domain with fish scale array was modeled in CATIA with a spanwise length equal to two wavelengths of the array $2\lambda_Z = 100$mm as shown in Fig.\ref{fig:figure2}C. The leading edge of the fish scale array was placed at $X = 300$mm from the leading edge of the flat plate as in the experiments. The length of the domain in the streamwise direction was set to 1200mm and in the wall normal direction the domain was 200mm in length. The inlet was specified with a velocity of $U_\infty = 0.086$m/sec and the outlet was specified as a pressure outlet. Periodic boundary conditions were used in the spanwise direction and the top domain was specified as free shear boundary with zero normal velocity. The domain was dicretised with 2mm elements in the streamwise direction, 1.43mm in the spanwise direction. The first cell height in the wall normal direction was set at 0.035mm with inflation of 20 cells within 2mm to capture the near wall gradient and a total of 110 grid points were used to mesh the wall normal direction. The major part of the domain was discretised with a Cartesian, structured mesh except the volume with fish scale array which was meshed with both prism and tetrahedral elements.The total number of elements for Setting-1 is 7.8 million and Setting-2 is 8.5 million. Second order pressure and second order upwinding schemes were used for discretisation with a steady state solver which was used to compute the laminar flow through the domain. \\

\bibliography{library}

\section*{Acknowledgements}

The position of Professor Christoph Bruecker is co-funded by BAE SYSTEMS and the Royal Academy of Engineering (Research Chair No. RCSRF1617$\backslash$4$\backslash$11, which is gratefully acknowledged. The position of MSc Muthukumar Muthuramalingam was funded by the Deutsche Forschungsgemeinschaft in the DFG project BR~1494/32-1 and  Dr. Dominik Puckert was supported by DFG project RI~680/39-1. 

\section*{Author contributions statement}

All authors conceived the experiment(s),  M.M., D.P. and C.B. conducted the experiments, M.M. and D.P. analysed the results. Initial draft was prepared by M.M and D.P. Finalised version was prepared with the contribution from all authors. The manuscript was based on the ideas and discussion on the transition delay.

\section*{Additional information}

\textbf{Accession codes} (where applicable);\\ \textbf{Financial Competing interests} The authors declare no competing interests. \\ \textbf{Non-Financial Competing interests} The authors declare no competing interests.  
 
\end{document}